\documentclass[journal]{IEEEtran}
\usepackage{graphicx}
\usepackage{setspace}
\usepackage{amsthm}
\usepackage{tabularx}
\usepackage{arydshln}
\usepackage{amssymb}
\usepackage{amsfonts}
\usepackage{mathrsfs}
\usepackage{amssymb,amsmath}
\usepackage{epstopdf}
\usepackage{bm}
\usepackage{algorithm}
\usepackage{algorithmic}
\usepackage{amsmath,amssymb,epsfig,graphics,subfigure}
\usepackage{flushend}
\usepackage{theorem}
\usepackage{array,color}
\usepackage[compress]{cite}
\usepackage{stfloats}
\usepackage{graphicx}
\usepackage{subfigure}
\usepackage{flushend}
\usepackage{array}
\usepackage{cases}
\usepackage{multirow}
\usepackage{chngpage}
\usepackage{longtable}
\usepackage{lineno}
\usepackage{caption}
\usepackage{amsfonts}
\usepackage{mathrsfs}
\usepackage{amssymb,amsmath}
\usepackage{algorithm}
\usepackage{algorithmic}
\usepackage{amsmath,amssymb,epsfig,graphics,subfigure}
\usepackage{theorem}
\usepackage{array,color}

\usepackage{amssymb}

\theoremheaderfont{\normalfont\bfseries}

\begin{document}


\title{Designing Enhanced Multi-dimensional Constellations for Code-Domain NOMA}
\author{Haifeng~Wen,
        Zilong~Liu,
        Qu~Luo,
        Chuang~Shi,
        and~Pei~Xiao
\thanks{Haifeng Wen and Chuang Shi are with University of Electronic Science and Technology of China (email: wenhaifeng@std.uestc.edu.cn, shichuang@uestc.edu.cn). Z. Liu is with the School of Computer Science and Electronics Engineering, University of Essex, UK (email: zilong.liu@essex.ac.uk). Qu~Luo and~Pei~Xiao are with 5GIC \& 6GIC, Institute for Communication Systems (ICS), University of Surrey, UK (email: \{q.u.luo,pxiao\}@surrey.ac.uk)}}
 \maketitle

 \begin{abstract}
This paper presents an enhanced design of multi-dimensional (MD) constellations which play a pivotal role in many communication systems such as code-domain non-orthogonal multiple access (CD-NOMA). MD constellations are attractive as their structural properties, if properly designed,  lead to signal space diversity and hence improved error rate performance. Unlike the existing works which mostly focus on MD constellations with large minimum Euclidean distance (MED), we look for new MD constellations with \textit{additional} feature that the minimum product distance (MPD) is also large.  To this end, a non-convex optimization problem is formulated and then solved by the convex-concave procedure (CCCP).
Compared with the state-of-the-art literature, our proposed MD constellations\footnote{All the obtained MD constellations can be found in https://github.com/Aureliano1/Multi-dimensional-constellation} lead to significant error performance enhancement over Rayleigh fading channels whilst maintaining almost the same performance over the Gaussian channels. To demonstrate their application, we also show that these MD constellations give rise to good codebooks in sparse code multiple access systems.
\end{abstract}


\vspace{-1em}
\section{Introduction}
 A multi-dimensional (MD) constellation refers to a set of equal-length vectors that exhibits certain distance properties. At the transmitter, several incoming bits are grouped to select a vector from an MD constellation which is then sent out over a multi-channel communication system (e.g., a multicarrier system). A judiciously designed MD constellation yields a large constellation shaping gain owing to the so-called signal space diversity \cite{Boutros1996,Boutros1998}. In general, the performance of an MD constellation is measured by its minimum Euclidean distance (MED) and/or minimum product distance (MPD). Specifically, a large MED leads to reliable detection in a Gaussian channel, whereas a large MPD is preferred for robust transmissions in a Rayleigh fading channel.

In recent years, the search for good MD constellations has attracted significant research attention due to their contemporary application in code-domain non-orthogonal multiple access (CD-NOMA) systems \cite{Liu2021}. Widely regarded as an enabling paradigm for massive connectivity in future machine-type communication networks, the codebook design of a CD-NOMA system relies on certain good MD constellations. A representative CD-NOMA scheme is sparse code multiple access (SCMA), in which low-complexity message passing decoding is carried out at the receiver by efficiently exploiting the sparsity structure of the codebooks \cite{LDS2008,SCMA}. In SCMA,  certain user-specific operations (such as interleaving, permutation, shuffling, phase rotations) may be applied to a common MD constellation to generate multiple sparse codebooks \cite{SCMADesign}. A comprehensive survey on various MD constellations for uplink SCMA codebook design is given in \cite{Marsland2019}. In \cite{Yu2018}, Star-QAM based MD constellation with large MED has been proposed for downlink SCMA. Recently, such an MD constellation is employed to construct power-imbalanced SCMA codebooks in \cite{Liu2021b}. The applications of SCMA for massive access in 6G has also been discussed in \cite{Liu2021c}. Besides, in the case of multiple-input multiple-output (MIMO) transmission, there has been a large body of literature concerning achieving the channel capacity at a high signal-to-noise ratio (SNR) through certain MD constellations with large MED \cite{Zheng2002,Gohary2009,Tahir2019}.

The primary objective of this paper is to design enhanced MD constellations with both large MED and MPD.
Despite numerous research attempts, this class of MD constellations is rarely known in the literature, to the best of our knowledge. From the numerical optimization point of view, a remarkable algorithm (perhaps the only known algorithm so far) for good MD constellations has been introduced in \cite{Beko2012} by minimizing the total constellation energy subject to an MED constraint. As shown in Section IV, the MD constellations from \cite{Beko2012} may suffer from small MPD, making the resultant communication system highly susceptible to transmission errors in Rayleigh fading channels. 

With the aid of the inequality of arithmetic and geometric means, we first observe that the MPD of an MD constellation tends to grow if the element-wise distances of any two MD vectors are enlarged \cite{Liu2021}. Thus, our optimization problem is transformed to achieving the minimum total constellation energy, while at the same time, the maximum element-wise distances and a large MED value, which is balanced by a trade-off factor. Such a problem is then tackled by the convex- concave procedure (CCCP) \cite{CCCP} to obtain a sub-optimal solution. Our numerical simulation results show that the obtained MD constellations lead to significant error performance enhancement over Rayleigh fading channels while maintaining almost the same performance over the Gaussian channels as compared to that of \cite{Beko2012}. Additionally, by applying these new MD constellations, we obtain improved SCMA codebooks whose BER performances in uplink Rayleigh fading channels outperform (or comparable to) those representative ones in the literature.

\textsl{Notations:} $x, \mathbf{x}$ and $\mathbf{X}$ denote scalar, vector and matrix, respectively. The $n$-dimensional complex and binary vector spaces are denoted as $\mathbb{C}^n$ and $\mathbb{B}^n$, respectively. Similarly, $\mathbb{C}^{k\times n}$ and $\mathbb{B}^{k\times n}$ denote the $(k\times n)$-dimensional complex and binary matrix spaces, respectively. $\text{tr}(\mathbf{X})$ denotes the trace of a square matrix $\mathbf{X}$. $\text{diag}(\mathbf{x})$ gives a diagonal matrix with the diagonal vector of $\mathbf{x}$. $(\cdot)^T$ and $(\cdot)^H$ denote the transpose and the Hermitian transpose. $\text{vec}(\cdot)$ denotes the vectorization operator. $\|\mathbf{x}\|_2$ and $|x|$ return the Euclidean norm of vector $\mathbf{x}$ and the absolute value of $x$, respectively.  $\mathbf{I}_N$ and $\mathbf{0}_N$ denote the identity matrix of order $N$ and the all zero matrix with size $N\times N$, respectively.

\vspace{-0.5em}
\section{Preliminaries}
\subsection{MD constellation design}
We consider a scenario where the transmit bits are mapped to a $K$-dimensional complex constellation with cardinality of $M$.
An MD constellation is denoted by $\mathcal {C}^{K\times M}=[\mathbf{x}_1,\mathbf{x}_2,...,\mathbf{x}_M]$, where $\mathbf{x}_i\in \mathbb{C}^{K\times 1}$, $i=1,2,...,M$.
The key performance indicators (KPIs) for an MD constellation include MED, MPD and kissing number. The kissing number refers to the number of constellation pairs that have the same MED (or MPD) \cite{Marsland2019}. Since minimizing the kissing number is intractable, it is desirable to maximize the MED or MPD first as they are the dominating factors of the BER performances \cite{Marsland2019,Boutros1998}. Next, we introduce MED and MPD for an MD constellation.

\textsl{ 1)  MED}: The MED of an MD constellation  is defined as
\begin{equation}
\small
  d_{E,\text{min}}=\min \{ \|\mathbf{x}_i-\mathbf{x}_j\|_2,  1\le i < j \le M \}.
\end{equation}

\textsl{2) MPD}:  The  product distance (PD) between two $K$-dimensional complex constellation vectors,  $\mathbf{x}_i$ and $\mathbf{x}_j$, is defined as:
\begin{equation}
\small
  d_{P,i,j}= \prod_{{k \in \mathcal{K}_{ij}}}{|x_{i,k}-x_{j,k}|},
\end{equation}
where $x_{i,k}$ and $x_{j,k}$ are the $k$-th complex element of $\mathbf{x}_i$ and $\mathbf{x}_j$, respectively.  $\mathcal{K}_{ij}$ denotes the set of admissible $k$, for which $x_{i,k} \neq  x_{j,k}$. Then the MPD of such an MD constellation is given by
 \begin{equation}
 \small
   d_{P,\text{min}}=\min \{  d_{P,i,j}, 1\le i < j \le M \}.
 \end{equation}

Maximizing the MED and MPD  of an MD constellation are key for the reliable transmission over Gaussian and Rayleigh fading channels, respectively \cite{Boutros1996,Boutros1998}. Given a power budget with the goal of maximizing both MED and MPD, the design problem can be formulated as
 \begin{equation} \label{eq_opt}
 \small
\begin{aligned}
  \begin{split}
     \max_{\mathcal{C}} & \quad \{d_{E,\text{min}},d_{P,\text{min}}\}  \\
     s.t. & \quad \frac{1}{M} \text{tr}(\mathcal{C}^H\mathcal{C})  = P,
  \end{split}
\end{aligned}
 \end{equation}
where $P$ is the average power of the constellation vectors. 
It is worth mentioning that the MED and MPD are also two design KPIs  for the CD-NOMA system \cite{Marsland2019}.

\vspace{-1em}
 \subsection{Introduction to SCMA}

To illustrate the application of the aimed MD constellations, we first provide a brief introduction to SCMA. Let us consider an  SCMA system with $J$ users communicating over  $N$ orthogonal resource nodes. To enable massive connectivity, the number of users is normally larger than that of resources, i.e. $J > N$ and thus the overloading factor is defined as $ \lambda_f = \frac{J}{N}> 1 $.   At the transmitter, for the $j$-th  user, the SCMA encoder maps $\log_2\left(M\right)$ coded binary bits to a $K$-dimensional complex codebook set $\mathcal {X}_{j}$, which is defined as
 $f_j:\mathbb{B}^{\log_2M}  \rightarrow {\mathcal X}_{j}   \in \mathbb {C}^{K}$, where  $\mathcal {X}_{j}=\left[\mathbf {x}_{j,1}, \mathbf {x}_{j,2},\ldots,\mathbf {x}_{j,M}\right] $   is the codebook of user $j$ with cardinality of $M$. All the $N$-dimensional complex codewords in each codebook are sparse vectors with $K$ non-zero elements and $K < N$.

 The design  of  optimal multi-dimensional codebooks for SCMA is still an open issue, hence a sub-optimal multi-stage design is generally considered  \cite{Yu2018}.
Let us consider  $\mathbf {V}_{j} \in \mathbb {B}^{N \times K} $ which is a mapping matrix associated with user $j$ that maps the $K$-dimensional constellation point to an $N$-dimensional sparse SCMA codeword.   The mapping matrix $\mathbf {V}_j$ is designed with $N-K$ all-zero rows, i.e., the all-zero  elements in $\boldsymbol { \mathcal X}_{j}$ are in the same dimensions with $\mathbf {V}_j$. The structure of SCMA codebook $\mathcal {X}$  can be represented by an indicator matrix  $\mathbf {F} = \left [ \mathbf {f}_1, \ldots, \mathbf {f}_J \right]$, where  $\mathbf {f}_j = \text {diag}(\mathbf {V}_j\mathbf {V}_j^T)$. User $j$ and resource $n$ are connected if and only if $f_{n,j}=1$,  where  $1 \le n \le N$, and $1 \le j \le J$.

Based on $\mathbf{V}_j$, user $j$'s codebook  is generated by $\boldsymbol{\mathcal X}_{j} = \mathbf {V}_{j} \mathbf{\Delta}_j {\mathcal A}_{MC}$,  where $\mathbf{\Delta}_j$ is the constellation operator of user $j$, $1\le j\le J$ and ${\mathcal A}_{MC}$ is the  MD constellation which is to be optimized in the next section. Similar to \cite{Liu2021b}, we combine the constellation operation matrix $\mathbf{\Delta}_j$  and mapping matrix $\mathbf {V}_{j}$ together, i.e., $\mathbf{s}_{N \times J}^j =\mathbf {V}_{j} \mathbf{\Delta}_j \mathbf{I}_K$, where $\mathbf{I}_K$ denotes a column vector  of $K~ 1$'s. Hence, the codebook can be represented by the signature matrix  $\mathbf{S}_{N \times J}= \left[ \mathbf{s}_{N \times J}^1, \dots, \mathbf{s}_{N \times J}^J \right]$.




\section{Proposed Optimization Method}
In this section, an MD constellation optimization scheme for both large MED and MPD is proposed. Our idea is to transform the MPD and MED constraints into a sequence of quadratic forms with linear inequality constraints.

Since it is quite difficult, if not impossible, to directly solve the optimization problem in (\ref{eq_opt}), we consider a  feasible way by transforming it into a single target problem. An equivalent problem is to find the minimum energy constellation $\mathcal{C}$ while keeping the MPD and MED greater or equal to the   thresholds $D_P$ and $D_E$, respectively, i.e.,
\begin{align}\label{single}
\small
  \begin{split}
    \min_{\mathcal{C}} &  \quad \text{tr}(\mathcal{C}^H\mathcal{C})  \\
   s.t.  & \quad \begin{array}{lc}
      d_{E,\text{min}} \ge D_E, \\
      d_{P,\text{min}} \ge D_P.  \end{array}
  \end{split}
\end{align}
The detailed settings  of $D_E$ and $D_P$ will be discussed later. Unfortunately, the optimization problem in (\ref{single}) is still hard to solve due to the non-convex constraints of MED and MPD. We thus reformulate the expressions of Euclidean distance and product distance in linear inequality constraints and    quadratic forms.

\textit{ Reformulation of MED: } Define $\mathbf{c}= \text{vec}(\mathcal{C}) \in \mathbb{C}^{KM\times 1}$. Then,  the Euclidean distance square between $\mathbf{x}_i$ and $\mathbf{x}_j$ can  be expressed in the following quadratic form:
\begin{equation} \label{med}
\small
  d_{E,i,j}^2=\|\mathbf{x}_i-\mathbf{x}_j\|_2^2=\mathbf{c}^H\mathbf{E}_{i,j}\mathbf{c}, \quad i\ne j,
\end{equation}
where  $\mathbf{E}_{i,j}=\mathbf{E}_i^T\mathbf{E}_i - \mathbf{E}_i^T\mathbf{E}_j - \mathbf{E}_j^T\mathbf{E}_i + \mathbf{E}_j^T\mathbf{E}_j  $, $\mathbf{E}_i=\mathbf{e}_i^T\otimes \mathbf{I}_K, 1\le i < j \le M$, $\mathbf{e}_i$ represents the $i$-th column of the identity matrix $\mathbf{I}_M$, and $\otimes$ denotes the Kronecker product.
The matrix $\mathbf{E}_{i,j}$ is very sparse, and with nonzero entries limited to $-1$ and $1$. Let $+$ and $-$ be $+1$ and $-1$, respectively. For $K=2$ and $M=4$, as an example, $\mathbf{E}_{i=1,j=2}$ is given below:


\begin{equation}
    \mathbf{E}_{i=1,j=2}= \left[
    \begin{array}{{c|c}}
    \begin{smallmatrix}
    +  &   0  &   -  &   0 \\
    0  &   +  &   0  &   - \\
    -  &   0  &   +  &   0 \\
    0  &   -  &   0  &   + \\
  \end{smallmatrix} & \mathbf{0}_{4} \\ \hline
  \mathbf{0}_{4} & \mathbf{0}_{4}
    \end{array} \right].
\end{equation}


\textit{ Reformulation of MPD: }  To  derive the  quadratic form of product distance square, we first give the equivalent expression of product distance square in a logarithmic form, i.e.,
\begin{equation}
\small
 \text{log}(d^2_{P,i,j})=\sum_{k\in \mathcal{K}_{ij}}{\text{log}(d_{i,j,k}^2)},
\end{equation}
where $d_{i,j,k}^2=|x_{i,k}-x_{j,k}|^2$ is the element-wise distance square at the $k$-th dimension. The quadratic form of $d_{i,j,k}^2$ at the $k$-th dimension can be easily obtained by padding zeros to other  dimensions. Similar to (\ref{med}), we have
\begin{equation}
\small
d_{i,j,k}^2=\mathbf{c}^H\mathbf{B}_{i,j,k}\mathbf{c}, \quad i\ne j, k\in \mathcal{K}_{ij},
\end{equation}
where $\mathbf{B}_{i,j,k}$ is obtained by keeping the $((i-1)K+k)$ and $((j-1)K+k)$ columns of $\mathbf{E}_{i,j}$, while replacing all the remaining columns with zeros.
As an example of $\mathbf{B}_{i,j,k}$ for $K=2$ and $M=4$ is given below:

\begin{equation}
    \mathbf{B}_{i=1,j=2,k=1}= \left[
    \begin{array}{{c|c}}
  \begin{smallmatrix}
    +  &   0  &   -  &   0 \\
    0  &   0  &   0  &   0 \\
    -  &   0  &   +  &   0 \\
    0  &   0  &   0  &   0 \\
  \end{smallmatrix} & \mathbf{0}_{4} \\ \hline
  \mathbf{0}_{4} & \mathbf{0}_{4}
    \end{array} \right].
\end{equation}


Based on the above analysis, the optimization problem in (\ref{single}) can be reformulated as:
\begin{equation}
\small
\begin{aligned} \label{opt_1}
  \begin{split}
     \min_{\mathbf{c}, t} & \quad t  \\
    s.t. &  \quad
      \|\mathbf{c}\|_2 \le t, \\
      & \quad \mathbf{c}^H\mathbf{E}_{i,j}\mathbf{c} \ge D_E^2, \\
      & \quad \sum_{k\in \mathcal{K}_{ij}}{\text{log}(\mathbf{c}^H\mathbf{B}_{i,j,k}\mathbf{c})} \ge 2\text{log}(D_P), \\
      & \quad 1 \le i < j \le M.
  \end{split}
\end{aligned}
\end{equation}

Albeit the above problem can be solved  by  semi-definite  relaxation (SDR), the resultant solutions may suffer from small MED values. 
This is due to the fact that the product distance in logarithmic form belongs to exponential cone which is considerably hard to be solved by symmetric primal/dual solvers\cite{Boyd2004}, thus affecting the performance of solving the optimization problem in (\ref{opt_1}) and leading to MED degradation.
A possible solution to this problem is to further  relax the MPD constraints.
Liu and Yang have shown in  \cite{Liu2021} the relationship between Euclidean distance and product distance by utilizing the inequality of arithmetic and geometric means, which is given by
\begin{equation} \label{ineq}
\small
  \begin{split}
    & d_P^2=\prod_{k=1}^K|x_{i,k}-x_{j,k}|^2
     \quad \le \left(\frac{\sum_{k=1}^{K}|x_{i,k}-x_{j,k}|^2}{K} \right)^K  \\
    & \quad =\frac{1}{K^K}\|\mathbf{x}_i-\mathbf{x}_j\|_2^{2K},
  \end{split}
\end{equation}
where the equality is achieved if and only if
\begin{equation} \label{equa}
  |x_{i,1}-x_{j,1}|=|x_{i,2}-x_{j,2}|=\text{...}=|x_{i,K}-x_{j,K}|.
\end{equation}

Based on (\ref{ineq}) and (\ref{equa}), the strong product distance constraints can be relaxed by element-wise distance constraints. Let us consider that the element-wise distance is no less than a threshold $\delta$, i.e., $|x_{i,k}-x_{j,k}|\ge \delta, \forall i,j,k$. Then the  product distance $d_{P}$ is  lower bounded by $\delta^K$, i.e.,
\begin{equation}
\small
  d_{P}\ge \delta^K.
\end{equation}
With this relaxation, more degrees of freedom may be exploited for maximizing the Euclidean distance while maintaining  high  energy efficiency. Hence, the optimization problem in (\ref{opt_1}) can be translated to the following one:
\begin{align} \label{qua-opt}
\small
  \begin{split}
    &\min_{\mathbf{c},t,\eta} \,\, t-\lambda \eta \\
    &s.t.\quad  \begin{array}{lc}
    \|\mathbf{c}\|_2 \le t,   \\
    \mathbf{c}^H\mathbf{E}_{i,j}\mathbf{c} \ge D_E^2, \\
    \mathbf{c}^H\mathbf{B}_{i,j,k}\mathbf{c} \ge \eta, \\
    1 \le i < j \le M, k=1,2,...K,
   \end{array}
    \end{split}
\end{align}
where $\eta$ is the introduced auxiliary variable corresponding to $\delta^2$, and $\lambda > 0$ is a hyper-parameter which is used to strike a trade-off between the MED and MPD. Small $\lambda$ tends to give rise to large MED but small MPD, whereas large $\lambda$ leads to increased MPD but small MED, i.e. approaching the equality in (\ref{ineq}). Hence,  $\lambda$ needs to be fine-tuned. In Section IV, $\lambda$ is set to be $1/2$ such that the obtained MD constellations possess both large MED and MPD.
The aim of (\ref{qua-opt}) is to minimize the total constellation energy and maximize the element-wise distance while maintaining a large MED.
Although the optimization problem (\ref{qua-opt}) is still non-convex, it can be simply linearized. In this paper, such a problem is solved by convex-concave procedure (CCCP) \cite{CCCP}.

Specifically, the optimization problem (\ref{qua-opt}) can be solved by iteratively solving the following convex problem:
\begin{align} \label{SOCP}
\small
  \begin{split}
    &\min_{\mathbf{c},t,\eta} \,\, \quad t-\lambda \eta \\
    &s.t.\quad  \begin{array}{lc}
    \|\mathbf{c}\|_2 \le t,   \\
    \mathbf{c}_q^H \mathbf{E}_{i,j} \mathbf{c}+\mathbf{c}^H \mathbf{E}_{i,j} \mathbf{c}_q-\mathbf{c}_q^H \mathbf{E}_{i,j} \mathbf{c}_q\ge D_E^2, \\
    \mathbf{c}_q^H \mathbf{B}_{i,j,k} \mathbf{c}+\mathbf{c}^H \mathbf{B}_{i,j,k} \mathbf{c}_q-\mathbf{c}_q^H \mathbf{B}_{i,j,k} \mathbf{c}_{q}\ge \eta, \\
    1 \le i < j \le M, k=1,2,...K,
   \end{array}
    \end{split}
\end{align}
where the subscript $q$ indicates the index of iterations.

\begin{figure}
  \centering
   \setlength{\abovecaptionskip}{2pt}
  \includegraphics[width=3in]{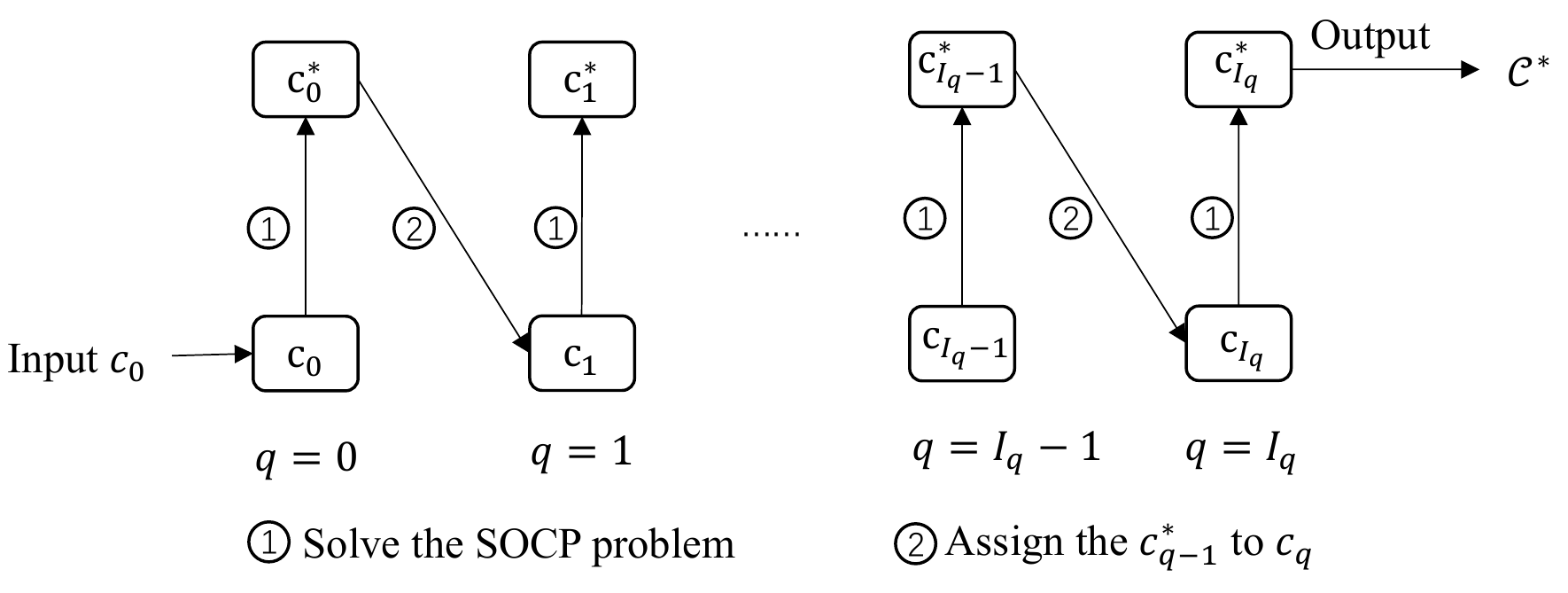}\\
  \caption{Illustration of the CCCP iterations in  solving the optimization problem (\ref{SOCP}). Note that the algorithm may stop when $\|\mathbf{c}_{q}-\mathbf{c}_{q-1}\|_2 \le \epsilon$.}
  \label{Diagram}
\end{figure}

It is noted that (\ref{SOCP}) is a convex second-order cone programming (SOCP) problem which can be solved by convex optimization tools, e.g. CVX.
The optimization process initiates with $q=0$ and a randomly sampled vector $\mathbf{c}_0$ that meets the constraints.  Then, in the $q$-th iteration, we solve (\ref{SOCP}) by  assigning  the  $(q-1)$-th optimized solution $\mathbf{c}_{q-1}^*$   to $\mathbf{c}_q$ to produce a new solution with a lower objective value.
Note that $\mathbf{c}^H\mathbf{E}_{i,j}\mathbf{c} \ge D_E^2$ and $\mathbf{c}^H\mathbf{B}_{i,j,k}\mathbf{c} \ge \eta$ are always satisfied during the iterations, since the left sides of the second and third constraints are their affine tight lower bounds.
Furthermore,  when $\mathbf{c}_{q-1}$ and $\mathbf{c}_{q}$ are feasible, $\|\mathbf{c}_{q}\|_2 \le \|\mathbf{c}_{q-1}\|_2$ is also satisfied, which means that the total constellation energy is not increasing during the iterations.
The algorithm stops when $\|\mathbf{c}_{q}-\mathbf{c}_{q-1}\|_2 \le \epsilon$ or the maximum number of iterations $I_q$ is reached. In our numerical simulation, $\epsilon$ and $I_q$  are   set to be $10^{-4}$ and $100$, respectively. Finally, the MD constellation   can be obtained by reshaping the latest optimized solution $\mathbf{c}^*$ to constellation matrix $\mathcal{C}^*$.
The above iteration process is illustrated in Fig. \ref{Diagram}.

According to the interior-point methods \cite{Boyd2004}, the worst-case computational complexity of the proposed method is $\mathcal{O}\left( \left( 1+(K+1)\left(\frac{M(M-1)}{2}\right)\right)^{3.5}\right)$,  which is  moderate even for large $M$ and $K$.  
\section{Numerical Evaluation}
\begin{figure}
  \centering
    \setlength{\abovecaptionskip}{2pt}
  \setlength{\belowcaptionskip}{-10pt}
  \includegraphics[width=3in]{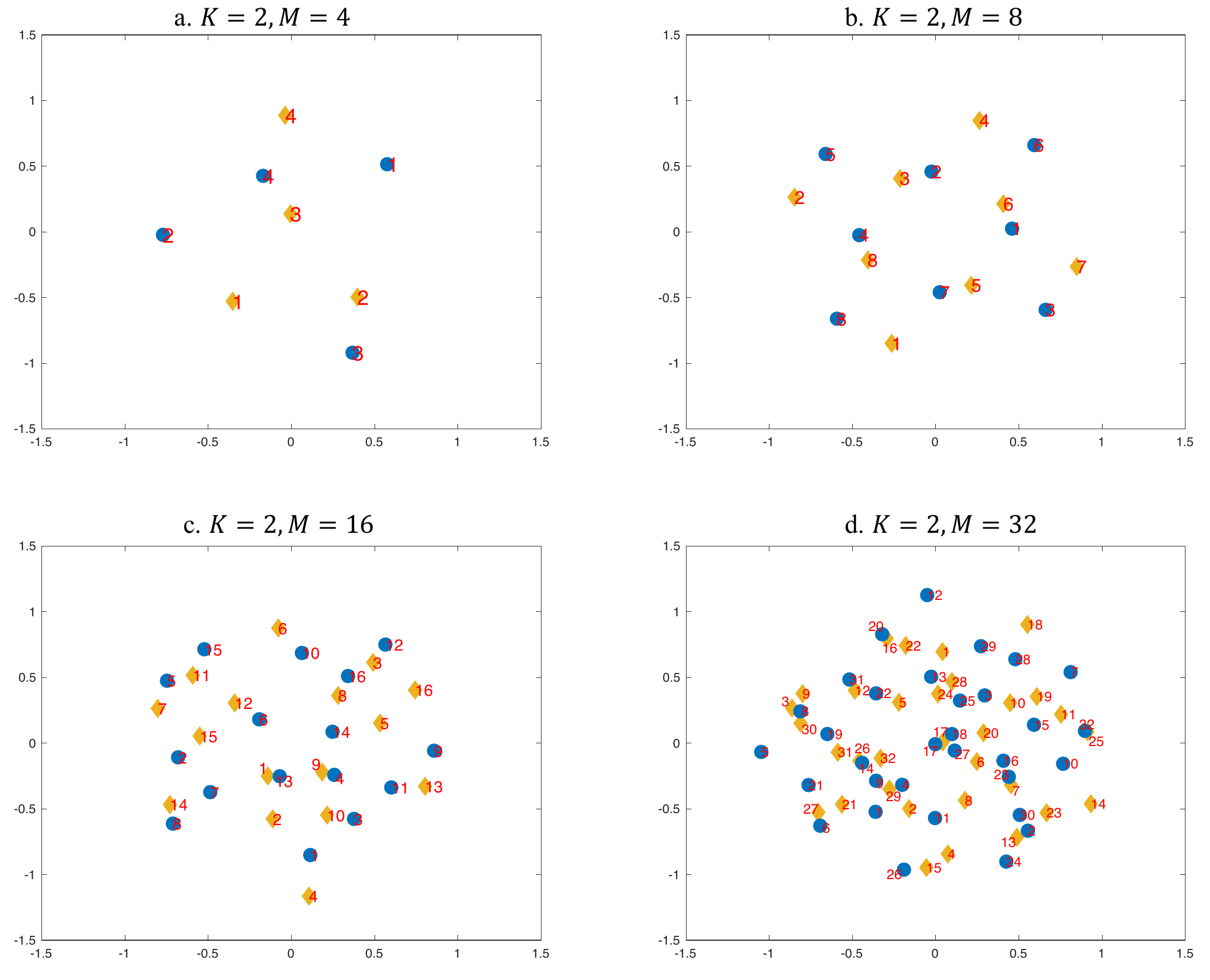}\\
  \caption{Plots of the proposed $ 2$-dimensional complex  constellations, where the blue circles refer to the first dimension, the yellow diamonds refer to the second dimension, and the index upon each circle/diamond stands for the corresponding MD vector number.}
  \label{2xMcons}
\end{figure}

This section presents a numerical evaluation of the proposed MD constellations. We first compare our proposed method with the   scheme in \cite{Beko2012} in terms of MED, MPD and BERs.
For all the simulations, $\lambda$ and $D_E$ are set to be 1/2 and 1, respectively.
The powers of all the obtained MD constellations are normalized to unit, i.e. $\text{tr}(\mathcal{C}^H\mathcal{C})/M=1$.
Based on these new MD constellations, we then construct $2$-dimensional SCMA codebooks for a $4\times 6$ SCMA system (which is widely adopted in the literature) and compare the error rate performances with some known SCMA codebooks in   uplink   Rayleigh fading channels. The indicator matrix of this $4\times 6$ SCMA system is given by
\begin{equation}
\small
   \setlength{\arraycolsep}{3.5pt}
\begin{aligned} \mathbf {S_{4 \times 6}}=\left [{
\begin{matrix}
0         &\quad 1 &\quad 1 &\quad 0 &\quad 1 &\quad 0 \\
1 &\quad 0          &\quad1 &\quad 0 &\quad 0 &\quad 1 \\
0         &\quad 1  &\quad 0 &\quad 1 &\quad 0 &\quad 1 \\
1 &\quad 0 &\quad 0 &\quad1 &\quad 1 &\quad 0 \\
\end{matrix} }\right].
\end{aligned}
\label{S_46}
\end{equation}


 Fig. \ref{2xMcons} presents the proposed $2$-dimensional constellations with  $M=4, 8,16,32$. As we can see, the constellation points at the same dimension, i.e., the  constellation points with the same color (same shape), own large MEDs. In other words, the element-wise distance at each dimension is optimized, thus helping contribute  a large MPD  for the obtained MD constellations. Due to the irregular pattern of constellation points, natural labeling is employed in the following BER simulations.



\begin{figure*}[htbp]
\footnotesize
\minipage{0.325\textwidth}
  \includegraphics[width=2.25in]{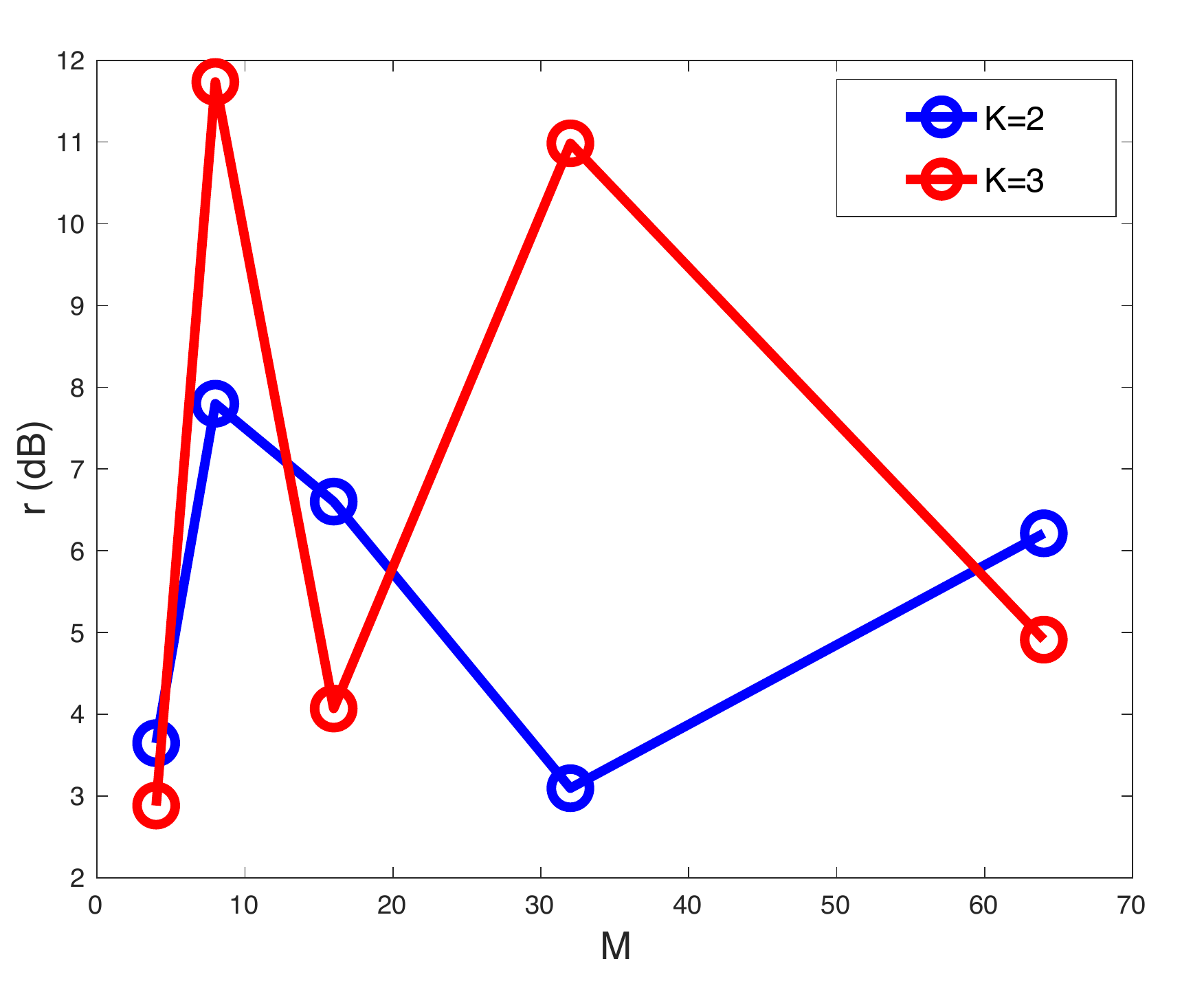}
  \caption{MPD improvement for the obtained MD constellations over that from  \cite{Beko2012} for $K = 2,3$.}
  \label{MPD_comparison}
				\vspace{-0.2em}
\endminipage\hfill
\minipage{0.325\textwidth}
  \includegraphics[width=2.25in]{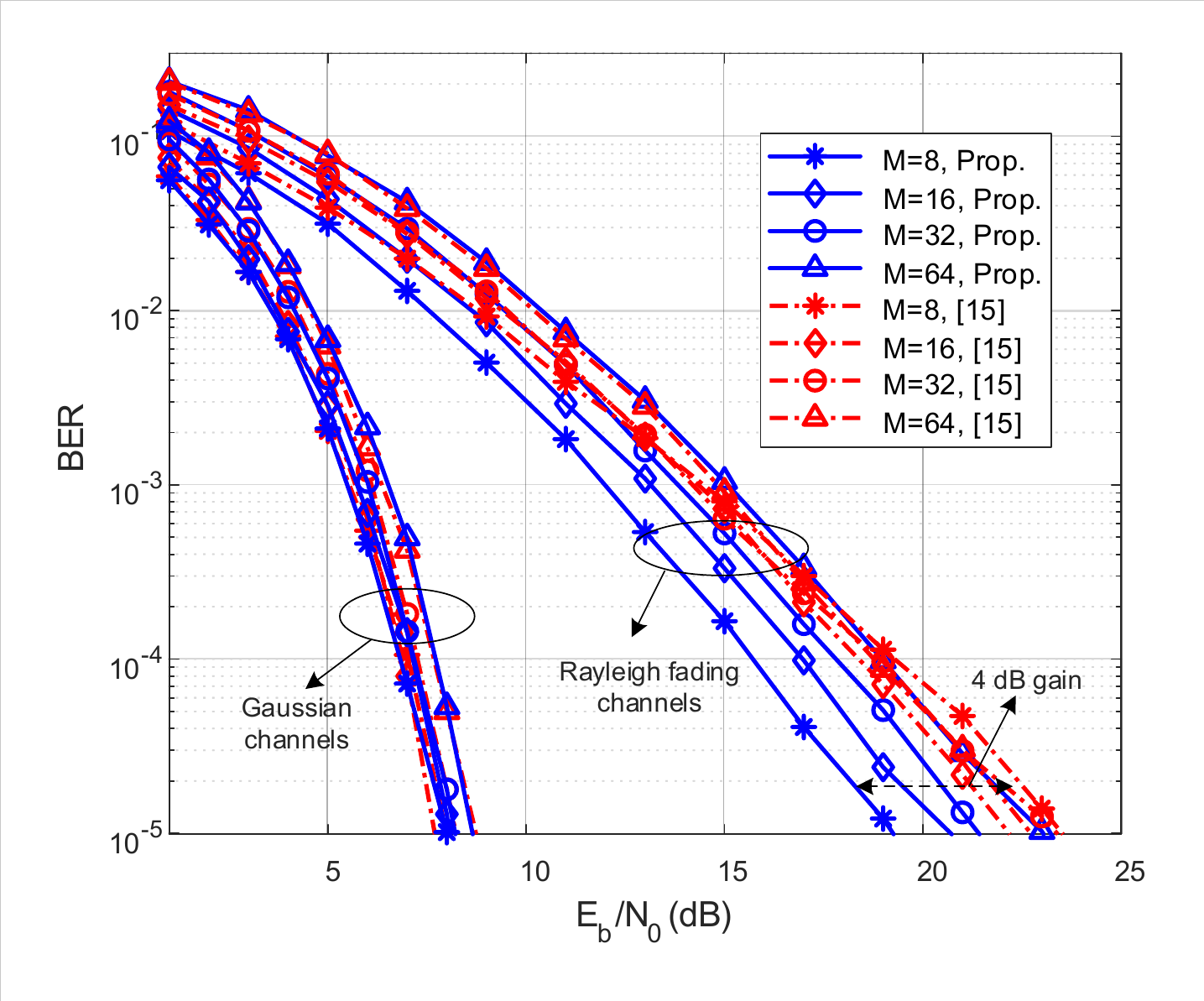}
\caption{Comparison of BERs between the obtained constellations and that from \cite{Beko2012} with $K=3$. }
 \label{BER_K=3}
				\vspace{-0.2em}
\endminipage\hfill
\minipage{0.325\textwidth}%
  \includegraphics[width=2.25in]{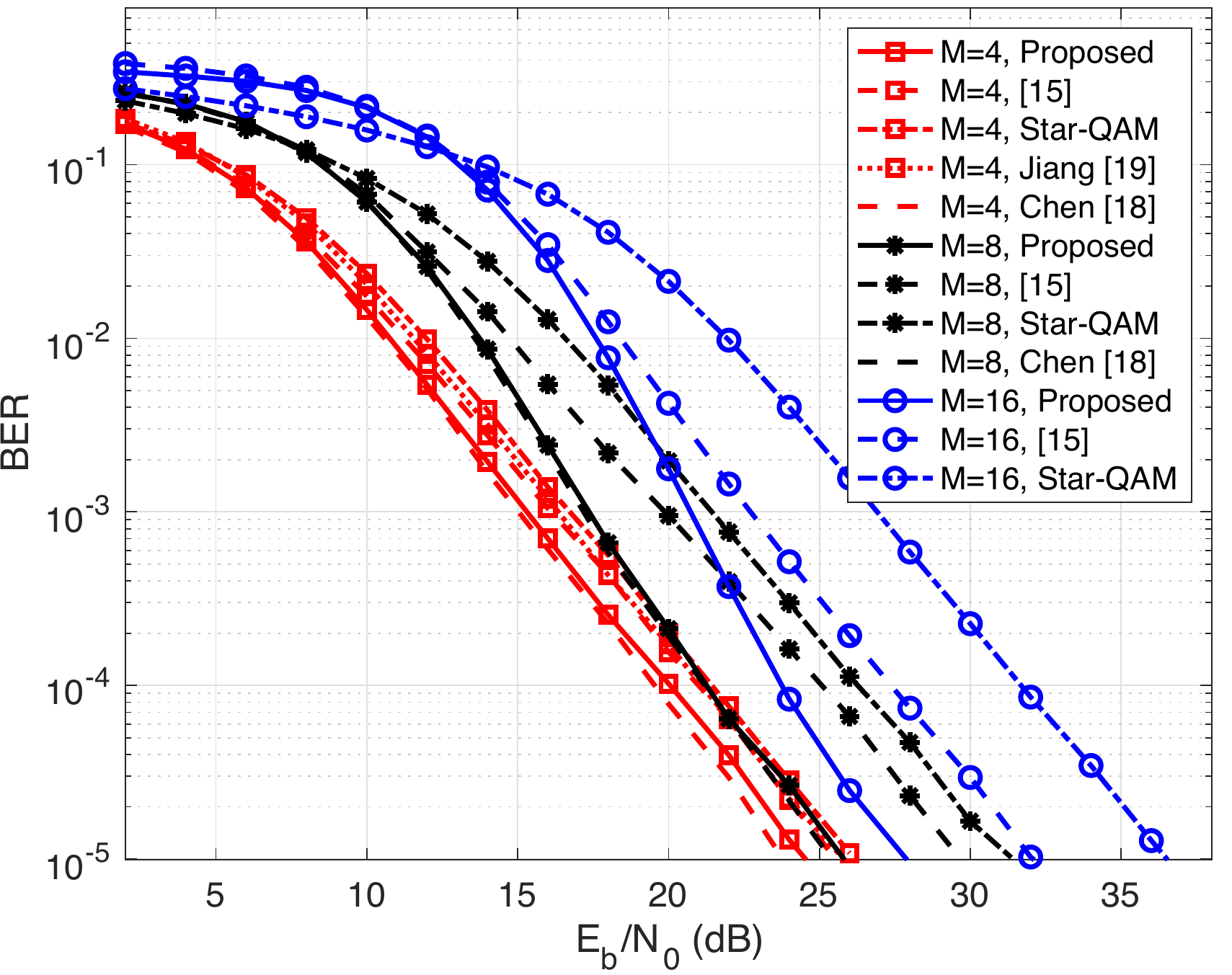}
  \caption{Comparison of BERs between different codebooks with $M=4, 8, 16$ in uplink SCMA under  Rayleigh fading channels.}
\label{up_scma_ber}
				\vspace{-0.2em}
\endminipage
				\vspace{-1.5em}
\end{figure*}

  Table \ref{MED_MPD_table} compares the MPD and MED values  of the obtained MD constellations with that from \cite{Beko2012} for $K \in \{2,3,5,7,9\}$ and $M\in \{4,8,16,32,64\}$. For the MD constellation with $K=2, M=4$, it has been proven in \cite{Huang2021} that the MED is upper bounded by $1.633$. We can observe that the obtained $2\times4$ constellation achieves the optimal MED  and large MPD values  at the same time. Moreover, the MPDs of the obtained MD constellations are significantly larger than those arising from  \cite{Beko2012}.  Let us denote by $r(K,M)$ the ratio between the MPDs of our obtained MD constellations with dimension of $K\times M$ and that from \cite{Beko2012}. Fig. \ref{MPD_comparison} shows the MPD improvements of the proposed method  over \cite{Beko2012} for $K=2,3$.
  It is noted that our obtained MD constellations outperform the MD constellations from \cite{Beko2012} by $3$ dB to $12$ dB for different $K$ and $M$. Since the MPD is not considered in \cite{Beko2012}, the MPD values of their MD constellations are random, thus leading to the  non-smooth ratio pattern.



\begin{table}[]
  \centering
  \scriptsize
\caption{A comparison of MED and MPD for the obtained MD constellations and that from \cite{Beko2012}, where the energy of the MD constellation is normalized to   $\text{tr}(\mathcal{C}^H\mathcal{C}) = M$.}
\begin{tabular}{|c|l|c|c||c|c|}
  \hline
  \multicolumn{2}{|c||}{\multirow{2}{*}{\textbf{$(K,M)$}}} & \multicolumn{2}{c||}{\textbf{Proposed}} & \multicolumn{2}{c|}{\textbf{\cite{Beko2012}}} \\ \cline{3-6}
  \multicolumn{2}{|c||}{}                                      & \textbf{MED}       & \textbf{MPD}      & \textbf{MED}           & \textbf{MPD}           \\ \hline
  \multicolumn{2}{|c||}{$(2,4)$}                            & 1.633              & 1.0887            & 1.633                  & 0.47                   \\ \hline
  \multicolumn{2}{|c||}{$(2,8)$}                            & 1.4142             & 0.8165            & 1.4142                 & 0.1355                 \\ \hline
  \multicolumn{2}{|c||}{$(2,16)$}                           & 1.1368             & 0.3572            & 1.127                  & 0.0781                 \\ \hline
  \multicolumn{2}{|c||}{$(2,32)$}                           & 0.9297             & 0.1166            & 0.9275                 & 0.0571                 \\ \hline
  \multicolumn{2}{|c||}{$(2,64)$}                           & 0.7599             & 0.0454            & 0.7531                 & 0.0109                 \\ \hline
  \multicolumn{2}{|c||}{$(3,4)$}                            & 1.633              & 0.7698            & 1.633                  & 0.3963                 \\ \hline
  \multicolumn{2}{|c||}{$(3,8)$}                            & 1.4759             & 0.3086            & 1.4771                 & 0.0207                 \\ \hline
  \multicolumn{2}{|c||}{$(3,16)$}                           & 1.2969             & 0.0906            & 1.3042                 & 0.0355                 \\ \hline
  \multicolumn{2}{|c||}{$(3,32)$}                           & 1.1415             & 0.0425            & 1.1495                 & 0.0034                 \\ \hline
  \multicolumn{2}{|c||}{$(3,64)$}                           & 0.9965             & 0.0114            & 0.9972                 & 0.0037                 \\ \hline
  \multicolumn{2}{|c||}{$(5,8)$}                            & 1.5119             & 0.05              & 1.5119                 & $1.45\times10^{-4}$               \\ \hline
  \multicolumn{2}{|c||}{$(5,16)$}                           & 1.4254             & 0.0158            & 1.4254                 & 0.004                  \\ \hline
  \multicolumn{2}{|c||}{$(7,8)$}                            & 1.5119             & 0.0043            & 1.5119                 & $4.45\times10^{-4}$              \\ \hline
  \multicolumn{2}{|c||}{$(7,16)$}                           & 1.4537             & $8.28\times10^{-4}$          & 1.4537                 & $1.38\times10^{-4}$              \\ \hline
  \multicolumn{2}{|c||}{$(9,8)$}                            & 1.5119             & $2.30\times10^{-4}$          & 1.5119                 & $3.72\times10^{-5}$              \\ \hline
  \multicolumn{2}{|c||}{$(9,16)$}                           & 1.4606             & $3.27\times10^{-5}$          & 1.4606                 & $6.82\times10^{-7}$               \\ \hline
  \end{tabular}

  \label{MED_MPD_table}
\end{table}


Then we compare the   BER performances of the obtained MD constellations and that from  \cite{Beko2012}  in Gaussian  and Rayleigh fading channels in  Fig. \ref{BER_K=3}. Specifically,  $K=3$ is considered and the maximum likelihood detector with perfect channel  information is assumed at the receiver side.  As can be seen, the proposed codebook achieve similar BER performance with that from \cite{Beko2012} in Gaussian channels, however, our obtained MD constellations outperform the constellations in \cite{Beko2012} significantly in Rayleigh fading channels in the high SNR region. One can  observe that our obtained MD constellations enjoy $4$ dB and $2$ dB gains for  $K=3, M=8$ and $  K=3, M=16$ respectively. To sum up, our proposed enhanced MD constellation can achieve good performance in both Gaussian and Rayleigh fading channels, due to the large MED and MPD.





By adopting the indicator matrix in (\ref{S_46}), Fig. \ref{up_scma_ber} compares the BER performances of the resultant SCMA codebooks (arising from the obtained MD constellations) with several representative codebooks with $K = 2$. These benchmarking codebooks for comparison are that obtained from [15], the Star-QAM codebook \cite{Yu2018}, Chen’s codebooks \cite{Chen2021} and Jiang’s codebook \cite{Jiang2022}.
Overall, our obtained codebooks outperform the Star-QAM codebooks, Jiang’s codebook and the codebooks from in \cite{Beko2012}, and achieve comparable (but slightly worse) BER performance with the codebooks in \cite{Chen2021} for $M = 4$ and $M = 8$.

The latter is because our obtained MD constellations exhibit relatively flat (or almost flat) MPD spectra, leading to limited error rate gain for optimizing the labeling between the input bits and transmit sparse codewords.  The excellent BER advantage of our obtained codebooks is more prominent for $M = 16$. In this case, the codebooks of \cite{Chen2021} are not available due to the intolerable computational complexity, whilst at the same time, we obtain about $5$ dB gain over that from \cite{Beko2012} and $10$ dB gain over Star-QAM codebooks at $\text{BER} = 10^{-5}$.






\section{Conclusions}

In this paper, we have proposed a new method of  designing enhanced MD constellations for CD-NOMA with large MED and MPD at the same time. The optimization problem is formulated (and then solved by CCCP) to minimize the energy of the MD constellations and maximize the element-wise distance while keeping the MED greater or equal to a certain threshold.
Whilst maintaining almost the same performance over the Gaussian
channels as compared to that of \cite{Beko2012}, numerical results have shown that the obtained MD constellations lead to enhanced error performances over Rayleigh fading channels, thanks to the enlarged MPD. Building upon these new MD constellations, good error performances are also observed for the resultant SCMA codebooks over uplink Rayleigh fading channels.


\end{document}